\documentclass[12pt,a4paper]{revtex4}
\usepackage{eurosym}
\usepackage[a4paper, total={6.5in, 9in}]{geometry}
\usepackage{float}
\usepackage{graphics}
\usepackage{graphicx}
\usepackage{amsmath}
\usepackage[title]{appendix}
\usepackage{subfigure}

\begin{document}

\title{Size effects in adhesive contacts of viscoelastic media}
\author{G. Violano}
\email{guido.violano@poliba.it}

\affiliation{Department of Mechanics, Mathematics and Management, Polytechnic University
of Bari, Via E. Orabona, 4, 70125, Bari, Italy}

\author{L. Afferrante}
\affiliation{Department of Mechanics, Mathematics and Management, Polytechnic University
of Bari, Via E. Orabona, 4, 70125, Bari, Italy}

\begin{abstract}
Is the maximum force required to detach a rigid sphere from a viscoelastic
substrate dependent on the initial value of the contact radius? Experimental
and theoretical investigations reported in the literature have given
opposite responses.

Here, we try to answer the above question by exploiting a fully
deterministic model in which adhesive interactions are described by
Lennard-Jones potential and the viscoelastic behaviour with the standard
linear solid model.

When the approach and retraction phases are performed under quasi-static
conditions, the substrate behaves as an elastic medium and, as expected, the
pull-off force $F_{\mathrm{PO}}$ (i.e., the maximum tensile force) is found
to be independent of the maximum contact radius $a_{\max }$ reached at the
end of loading. Size-dependent effects are instead observed (i.e., pull-off
force $F_{\mathrm{PO}}$ changes with $a_{\max }$) when transient effects occur as the larger the contact area, the greater the size of the
bulk volume involved in the dissipation. Results are also discussed in the
light of viscoelastic crack Persson's theory, which is modified to capture
size effects related to $a_{\max }$.
\end{abstract}
\maketitle
\section{Introduction}

The estimate of the pull-off force, required to detach two solids in
adhesive contact, is of crucial importance in several applications, such as
micro-transfer printing \cite{Meitl2006}, tactile sensors \cite{Yao2020},
biomimetic adhesives \cite{Mazzotta2020}, soft grippers \cite{Coulson2021}.
In all these technologies, there is a widespread use of soft compliant
materials, which often exhibit viscoelasticity and rate-dependent adhesive
features, which in turn lead to adhesion hysteresis in loading-unloading
cycle \cite{Violano2021a}. As a result, modeling adhesion between soft
matter is a tricky challenge, as shown by efforts in recent theoretical
studies \cite{Dokkum2021,Muser2021,Persson2021}.

We lately developed a fully deterministic model for the adhesive contact
between a rigid spherical indenter and a viscoelastic substrate \cite%
{Afferrante2021}. We observed that, when unloading starts from a fully
relaxed state of the substrate, the \textit{viscoelastic} pull-off force $F_{%
\mathrm{PO}}$, i.e., the maximun tensile force, is a monotonic increasing
function of the contact line velocity $V\mathrm{_{\mathrm{c}}}$ and reaches
an asymptotic value at high speeds. If we denote with $F_{\mathrm{PO,0}}$
the quasi-static pull-off force (i.e., its value in the elastic limit), at
large speeds the ratio $F_{\mathrm{PO}}/F_{\mathrm{PO,0}}$ should be
theoretically equal to $E_{\infty }/E_{0}$, being $E_{\infty }$\ and $E_{0}$
the high and low frequency viscoelastic moduli, respectively. However, $F_{%
\mathrm{PO}}/F_{\mathrm{PO,0}}$ is usually lower as a result of finite-size
effects. The trend of $F_{\mathrm{PO}}$ with $V\mathrm{_{\mathrm{c}}}$ is
completely different when unloading starts from an unrelaxed state of the
material as dissipation involves the bulk volume, and hence it is not
confined around the contact line \cite{Afferrante2021}.

Based on experimental evidence, researchers have given opposite answers
about the dependence of the pull-off force on the point from which unloading
starts. For example, Dorogin et al. \cite{Dorogin2017} carried out adhesion
tests between a spherical soda-lime glass ball indenter and a
PolyDiMethylSiloxane (PDMS) substrate. They used the same value for the
approach and retraction speeds of the ball, but they did not specify if any
dwell time was waited before unloading. In such conditions, they did not
observe a dependence of the pull-off force on the maximum preload and
explained that such behaviour is expected when the contact area is simply
connected, i.e., when the presence of surface roughness can be neglected.

Similarly, Violano et al. \cite{ViolAIAS2019} observed an almost constant
pull-off force in classical adhesion experiments between a spherical glass
indenter and a soft PDMS\ sample when unloading is performed from fully
relaxed state of the material and different preloads.

More recently, Das \& Chasiotis \cite{Das2021} performed experiments to
study rate-dependent adhesion between polymer nanofibers. To avoid
viscoelastic effects during the loading phase, a little crosshead velocity ($%
12$ \textrm{nm/s}) was fixed. Retraction was instead performed at greater
crosshead velocity ($2$ $\mathrm{\mu }$\textrm{m/s}). In such conditions,
they also observed $F_{\mathrm{PO}}$ to be independent of the preload.

Different results were instead obtained by Baek et al. \cite{Baek2017}, who
performed adhesion experiments between a PDMS\ block and a spherical glass
lens. To minimize viscous dissipation, lens approach was performed in
step-by-step movements (with a dwell time of $15$ \textrm{s}\ at the end of
each step). Retraction was performed at fixed speed until the detachment of
the lens. They found an increase in the pull-off force with $a_{\max }$ and,
hence, preload. They justified such effect as a consequence of the energy
dissipated at the contact interface, as theoretically explained by Maugis \&
Barquins (MB) \cite{MB1980}.

Kroner et al. \cite{Kroner2013} performed adhesion measurements between PDMS
samples and spherical probes. They observed a monotonous increase in the
pull-off force with increasing preload, with a maximum asymptotic value
reached for higher preloads. Similar results were also obtained in Ref. \cite%
{Lai2019}, where adhesion experiments were conducted on polyacrylamide
hydrogel.

From a theoretical point of view, the classical Johnson, Kendall \& Roberts
(JKR) theory predicts a constant pull-off force equal to $F_{\mathrm{PO,0}%
}=1.5\pi \Delta \gamma R$, being $R$ the radius of curvature of the sphere
and $\Delta \gamma $ the adiabatic surface energy. However, JKR theory \cite%
{JKR1971} assumes that contact occurs between elastic media, neglecting
viscoelasticity and rate-dependent adhesion.

Attard \cite{Attard2001} developed a numerical model for the adhesion of
viscoelastic spheres, observing that the pull-off force "\textit{will be
independent of the maximum applied load. The exception is when the maximum
applied load is relatively small}".

In a recent work, Jiang et al. \cite{Jiang2021} developed a finite element
model for the viscoelastic adhesive contact between a PDMS stamp and a
sphere. In their simulations, unloading starts from a relaxed state of the
viscoelastic stamp. They calculated higher pull-off force for increasing
preload. Such dependence was found to be more significant for larger
unloading velocities.

The detachment of a rigid sphere from a viscoelastic substrate is analogous
to the opening of a circular crack. In such case, the effective surface
energy $\Delta \gamma _{\mathrm{eff}}$ required to advance the crack tip by
one unit area increases with the crack tip velocity $V\mathrm{_{\mathrm{c}}}$
in the same way of the pull-off force \cite{Afferrante2021}. Two main
approaches have been formulated to study viscoelastic cracks, namely the
cohesive approach \cite{Schapery1975,Greenwood2004} and the energetic one 
\cite{Degennes,PB2005}, which yield very similar results (see Refs. \cite%
{Afferrante2021,Persson2021}).

Persson \cite{Persson2017} showed that, for system of finite size, $\Delta
\gamma _{\mathrm{eff}}$ does not reach an asymptotic value at high $V\mathrm{%
_{\mathrm{c}}}$, but it shows a maximum value at intermediate speeds. More
recently, Afferrante and Violano \cite{Afferrante2021} and the same Persson 
\cite{Persson2021} have shown that this is true only when unloading starts
from an unrelaxed state of the material. In the case of unloading from a
relaxed state, a different approach than Persson's one \cite{Persson2017}
needs to be conceived to capture finite-size effects related to $a_{\max }$.

Moving from this depicted state of the art, in the present paper, we try to
clarify how finite-size effects influence the pull-off force and, hence, the
effective surface energy $\Delta \gamma _{\mathrm{eff}}$. Specifically, we
investigate the influence of the initial contact size $a_{\max }$, from
which unloading starts, and propose a modification of Persson's theory for
viscoelastic crack to capture size-effects related to $a_{\max }$.

\section{Statement of the problem}

\begin{figure}[tbp]
\begin{center}
\includegraphics[width=8.0cm]{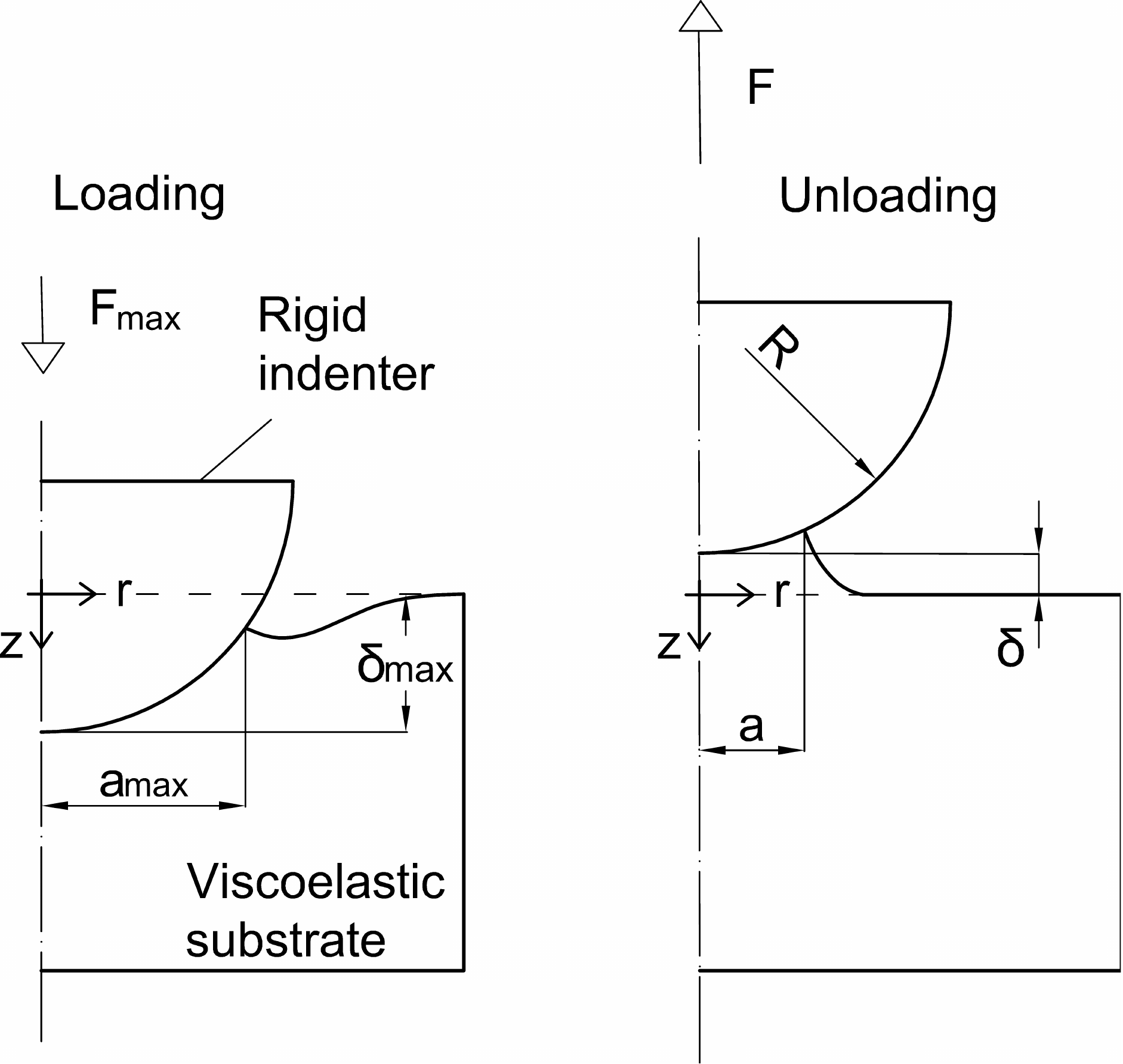}
\end{center}
\caption{The problem under investigation: a rigid sphere is pressed into a
viscoelastic substrate up to a maximum penetration $\protect\delta _{\max }$%
, which corresponds to a maximum contact radius $a_{\max }$ and preload $%
F_{\max }$. The sphere is then detached; both loading and unloading phases
are performed at fixed velocity $V=d\protect\delta /dt$.}
\label{FIGURE1}
\end{figure}

The problem under investigation is sketched in Fig. \ref{FIGURE1}: a rigid
spherical indenter (with radius of curvature $R$) is pressed into a
viscoelastic substrate until a fixed penetration $\delta $ is reached.
Different loading histories are considered by increasing the maximum
penetration $\delta _{\max }$ (and hence the maximum contact radius $a_{\max
}$); the sphere is then detached from the substrate.

\begin{figure}[tbp]
\begin{center}
\includegraphics[width=16.0cm,trim={0.5cm 0.8cm 0.5cm 0.8cm},clip]{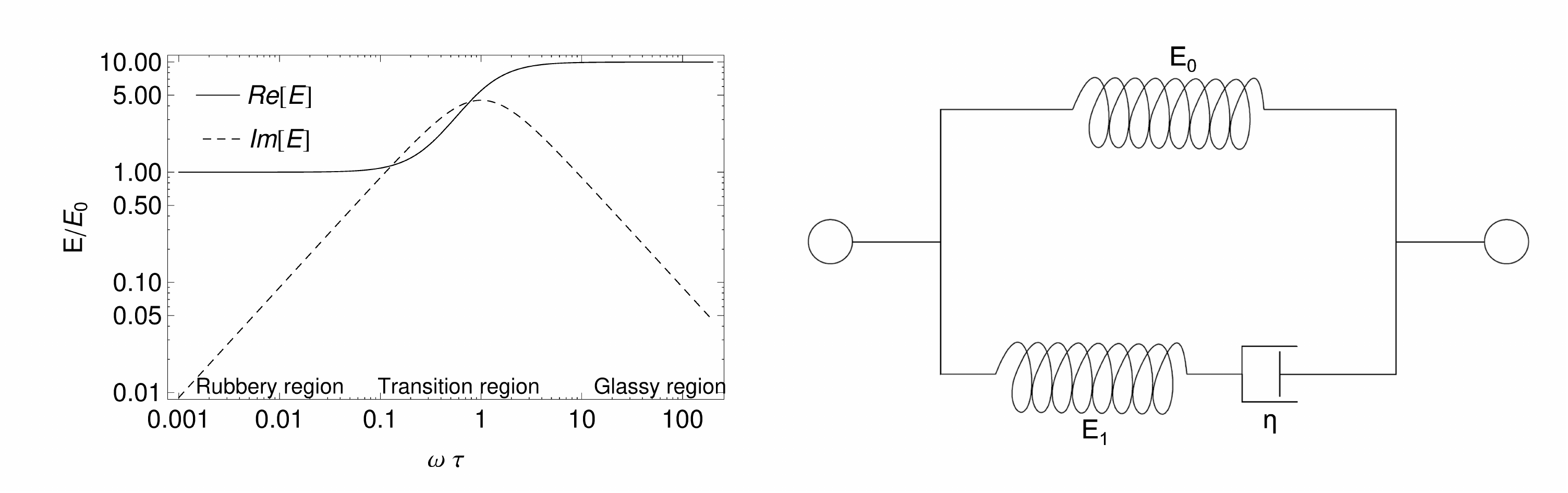}
\end{center}
\caption{Left: the real (solid line) and imaginary (dashed line) parts of
the viscoelastic modulus for a standard linear solid. Right: the standard
linear solid used in modelling the viscoelastic properties of the substrate. 
$E_{0}$ is the low frequency Young's modulus, $E_{1}=E_{\infty }-E_{0}$,
where $E_{\infty }$ is the high frequency Young's modulus, and $\protect\eta %
=E_{1}/\protect\tau $ is the viscosity of the dashpot, being $\protect\tau $
the relaxation time of the material.}
\label{FIGURE2}
\end{figure}

The constitutive behaviour of the substrate is modeled with a standard
linear solid, whose viscoelastic modulus $E$ is dependent on frequency $%
\omega $ as shown in Fig. \ref{FIGURE2}. At low frequencies, i.e., in the
rubbery region, the material behaves as a soft elastic medium with constant $%
E(\omega )=E_{0}$. At high frequencies, i.e., in the glassy region, a
stiffer elastic modulus $E(\omega )=E_{\infty }$ is experienced. At
intermediate frequencies, i.e., in the transition region, viscous
dissipation occurs and the material behaves as a viscoelastic medium.

\subsection{Numerical solution}

To solve the contact problem shown in Fig. \ref{FIGURE1}, we make use of the
deterministic finite element (FE) model developed in Ref. \cite%
{Afferrante2021}, at which the reader is referred for the details of the
formulation. Here, we briefly recall that interface interactions are modeled
by a traction-gap relation based on Lennard-Jones potential law according to
Derjaguin's approximation \cite{Derjaguin1934}. Moreover, Maxwell
representation (Fig. \ref{FIGURE2}B) of the classical linear standard model
is used to describe the viscoelastic behaviour of the substrate 
\begin{equation}
E\left( t\right) =E_{0}+\left( E_{\infty }-E_{0}\right) \exp \left( -t/\tau
\right)
\end{equation}%
where $\tau $ is the relaxation time.

Finally, the stress $\mathbf{\sigma }$ is calculated according to%
\begin{equation}
\mathbf{\sigma =}\int_{0}^{t}E\left( t-t^{\prime }\right) \frac{d\mathbf{%
\varepsilon }}{dt^{\prime }}dt^{\prime }  \label{s}
\end{equation}%
where $\mathbf{\varepsilon }$ is the strain and $E\left( t\right) $ is the
relaxation function.

\subsection{Modified Persson's theory}

\subsubsection{Unloading from relaxed state}

In the problem sketched in Fig. \ref{FIGURE1}, the contact mechanical
response is concurrently affected by the time-dependent behaviour of the
viscoelastic material and finite-size effects. Let us consider a first
scenario, where the sphere approaches the substrate at vanishing normal
speed $V=d\delta /dt\approx 0$. As a result, quasi-static conditions occur
and rate-dependent effects are negligible. In such case, the substrate is in
a fully relaxed state at the end of the loading phase. Notice the same
condition can be reached by loading at nonzero velocity and waiting a
sufficient long dwell time ($t_{0}\rightarrow \infty $) before unloading 
\cite{ViolAIAS2019,Jiang2021}. If unloading is then performed at nonzero $V$%
, the maximum pull-off force $F_{\mathrm{PO(\max )}}$, which is
asymptotically reached at high contact line velocities $V\mathrm{_{\mathrm{c}%
}}=-da/dt$ \cite{Afferrante2021}, should be 
\begin{equation}
\frac{F_{\mathrm{PO(\max )}}}{F_{\mathrm{PO,0}}}=\frac{E_{\infty }}{E_{0}}%
\text{,}  \label{maxFPO}
\end{equation}%
being $F_{\mathrm{PO,0}}$ the quasi-static pull-off force. However, the
ratio $F_{\mathrm{PO(\max )}}/F_{\mathrm{PO,0}}$ is usually lower than that
of eq. (\ref{maxFPO}).

In Ref. \cite{Ciava2021}, it is suggested that (\ref{maxFPO}) is true only
when adhesion is characterized by short-range adhesive interactions (Tabor
parameter \cite{Tabor} $\mu (\omega )=R^{1/3}\left[ \Delta \gamma (1-\nu
^{2})/E(\omega )\right] ^{2/3}\gg 5$). However, we suspect that geometric
and finite-size effects can also influence the pull-off force and, hence,
the effective surface energy $\Delta \gamma _{\mathrm{eff}}$ independently
of the value of $\mu $.

The detachment of a rigid sphere from a viscoelastic substrate is analogous
to the opening of a circular crack \cite{Persson2017}; thus, according to
Ref. \cite{Afferrante2021}%
\begin{equation}
\frac{\Delta \gamma _{\mathrm{eff}}}{\Delta \gamma }=\frac{F_{\mathrm{PO}}}{%
F_{\mathrm{PO,0}}}\text{.}
\end{equation}
Persson \& Brener (PB) \cite{PB2005} showed that the surface energy required
to advance the crack tip by one unit area is related to the crack tip radius 
$s(V_{\mathrm{c}})$ by%
\begin{equation}
\frac{\Delta \gamma _{\mathrm{eff}}}{\Delta \gamma }=\frac{s}{s_{0}}\text{,}
\label{PB2}
\end{equation}%
where $s_{0}$ is the crack tip radius for $V_{\mathrm{c}}\sim 0$ and is
related to the stress $\sigma _{\mathrm{c}}$ needed to break atomic bonds
through%
\begin{equation}
\sigma _{\mathrm{c}}=\frac{K_{\mathrm{I}}}{\left( 2\pi s_{0}\right) ^{1/2}}%
=\left( \frac{E_{0}\Delta \gamma }{2\pi s_{0}}\right) ^{1/2},
\end{equation}%
being $K_{\mathrm{I}}$ the stress intensity factor for mode \textrm{I}.

Moreover, from the energy conservation applied to the crack propagation, PB
extracted an equation relating the effective surface energy $\Delta \gamma _{%
\mathrm{eff}}$ (and, hence, the crack tip radius $s$ by eq. (\ref{PB2})) to
the viscoelastic modulus $E(\omega )$%
\begin{equation}
\Delta\gamma_{\mathrm{eff}}=\Delta\gamma\left[ 1-\frac{\frac{2E_{\infty}}{\pi}\int_{0}^{2\pi V_{\mathrm{c}}/s}\frac{F(\omega)}{\omega}Im\left( \frac{1}{E(\omega)}\right) d\omega }{1+\frac{2E_{\infty}}{\pi}\int_{0}^{2\pi V_{\mathrm{c}}/s}\frac{1}{\omega}Im\left(\frac{1}{E(\omega)}\right)d\omega}\right]^{-1}
\label{PB}
\end{equation}%
where $F(\omega )=[1-(\omega s/(2\pi V_{\mathrm{c}}))^{2}]^{1/2}$. Equation (%
\ref{PB}) leads to the identity (\ref{maxFPO}) at high $V_{\mathrm{c}}$, but
it can be used only under the assumptions of: (i) unloading started from a
relaxed state of the viscoelastic material, and (ii) system of infinite size.

To take account of the geometry and finite dimension of the contact area, we
observe that the problem of detachment of a rigid sphere from a flat
substrate, as shown in Fig. \ref{FIGURE1},\textbf{\ }is analogous to the
problem of a round shaft of radius $a_{\max }$ subjected to a tensile axial
load $F$ and with a circumferential crack of initial size $s_{0}$. In this
case, the stress $\sigma _{\mathrm{c}}$ at the tip of the circumferential
crack can be corrected according to $\sigma _{\mathrm{c}}=K_{\mathrm{I}}/%
\left[ \left( 2\pi s_{0}\right) ^{1/2}f(\beta )\right] $, being $f(\beta
)=0.5\beta ^{-1.5}(1+0.5\beta +0.375\beta ^{2}-0.363\beta ^{3}+0.731\beta
^{4})$ and $\beta =(1-s_{0}/a_{\max })$ (see Ref. \cite{Dowling}). As a
result, one can easily show that (\ref{PB}) modifies in%
\begin{equation}
\Delta \gamma _{\mathrm{eff}}=\Delta \gamma \left[ 1-\alpha \frac{\frac{%
2E_{\infty }}{\pi }\int_{0}^{2\pi V_{\mathrm{c}}/s}\frac{F(\omega )}{\omega }%
Im\left( \frac{1}{E(\omega )}\right) d\omega }{1+\frac{2E_{\infty }}{%
\pi }\int_{0}^{2\pi V_{\mathrm{c}}/s}\frac{1}{\omega }Im\left( \frac{1%
}{E(\omega )}\right) d\omega }\right] ^{-1}\text{,}  \label{PBnew}
\end{equation}%
where the parameter $\alpha $ depends on $s_{0}$ and $a_{\max }$, and is
given by%
\begin{equation}
\alpha =\left[ \frac{f(\beta )}{1.1215}\right] ^{-2}\text{.}  \label{alpha}
\end{equation}%
For vanishing $s_{0}/a_{\max }$, $f(\beta )\rightarrow 1.1215$, which is the
well-known result for edge cracks. In eq. (\ref{alpha}), the constant $%
1.1215 $ is hence introduced to obtain the same result of the original eq. (%
\ref{PB}) in the case of small $s_{0}/a_{\max }$ (namely, for very large
contact radius $a_{\max }$).

\subsubsection{Unloading from unrelaxed state}

When the loading-unloading cycle is performed at fixed nonzero velocity $%
V=d\delta /dt$ and unloading starts right after the loading phase, the
substrate material does not have the time to `relax'. Afferrante \& Violano 
\cite{Afferrante2021} showed that the pull-off force reaches a maximum value
at intermediate $V\mathrm{_{\mathrm{c}}}$ and then decreases by increasing
the contact line velocity. This behaviour has been observed also
experimentally in Ref. \cite{Luengo} and is related to the finite dimension
of the system under investigation \cite{Persson2021}.

Persson extended his theory of crack's propagation to the case of
finite-sized viscoelastic solids, with application to spheres adhesion \cite%
{Persson2017}, by introducing in the integrals of eq. (\ref{PB}) a cut-off
frequency $\omega _{\mathrm{L}}$ related to the dimension $L$ of the system%
\begin{equation}
\Delta \gamma _{\mathrm{eff}}=\Delta \gamma \left[ 1-\frac{\frac{2E_{\infty }%
}{\pi }\int_{\omega _{\mathrm{L}}}^{2\pi V_{\mathrm{c}}/s}\frac{F(\omega )}{%
\omega }Im\left( \frac{1}{E(\omega )}\right) d\omega }{1+\frac{%
2E_{\infty }}{\pi }\int_{\omega _{\mathrm{L}}}^{2\pi V_{\mathrm{c}}/s}\frac{1%
}{\omega }Im\left( \frac{1}{E(\omega )}\right) d\omega }\right] ^{-1}%
\text{.}  \label{PBnew2}
\end{equation}

We have shown, in Ref. \cite{Afferrante2021}, that viscous
dissipation is no longer confined at the contact edge, but may involve
bulk material when unloading starts from an unrelaxed state. In this case,
according to Ref. \cite{Persson2017}, we expect that size effects are governed by the spectrum
of frequencies $\omega $ considered in eq. (\ref{PBnew2}). As a result,
the parameter $\alpha $ must not be considered here, as
its derivation assumes detachment is governed by local effects around the crack tip.

We stress that eq. (\ref{PBnew2}) works only when retraction of the sphere
starts from an unrelaxed state of the material and the relaxation modes are not able to totally recover their undeformed state \cite{Persson2021}. In this case, it is reasonable to exclude the frequencies $\omega <\omega _{\mathrm{L}}$, being the cut-off dimension $L$ of the order of $a_{\max}$ \cite{Persson2017}.

\section{Results}

All results are obtained for $E_{\infty }/E_{0}=10$, Poisson's ratio $\nu
\approx 0.5$, and are given in terms of dimensionless quantities:\ $\hat{R}%
=R/\varepsilon $, $\hat{V}=V\tau /\varepsilon $, $\hat{a}=a/\varepsilon $, $%
\hat{\delta}=\delta /\varepsilon $, $\hat{F}=F/(1.5\pi \Delta \gamma R)$,
being $\tau $ the relaxation time of the viscoelastic material and $%
\varepsilon $ the range of action of van der Waals forces. All simulations
are performed under displacement controlled conditions and for fixed normal
velocity $V$ of the spherical indenter.

\subsection{Unloading from relaxed state}
A first set of simulations has been run at vanishing approaching velocity ($%
V\approx 0$) to avoid time dependent effects during the loading phase and to
ensure the detachment process starts from a complete relaxed state of the
substrate. Unloading is instead performed at different speeds, starting from
different maximum contact penetrations $\delta_{\max }$ (and, hence,
different values of $a_{\max }$).

\begin{figure}[tbp]
\begin{center}
\includegraphics[width=16.0cm]{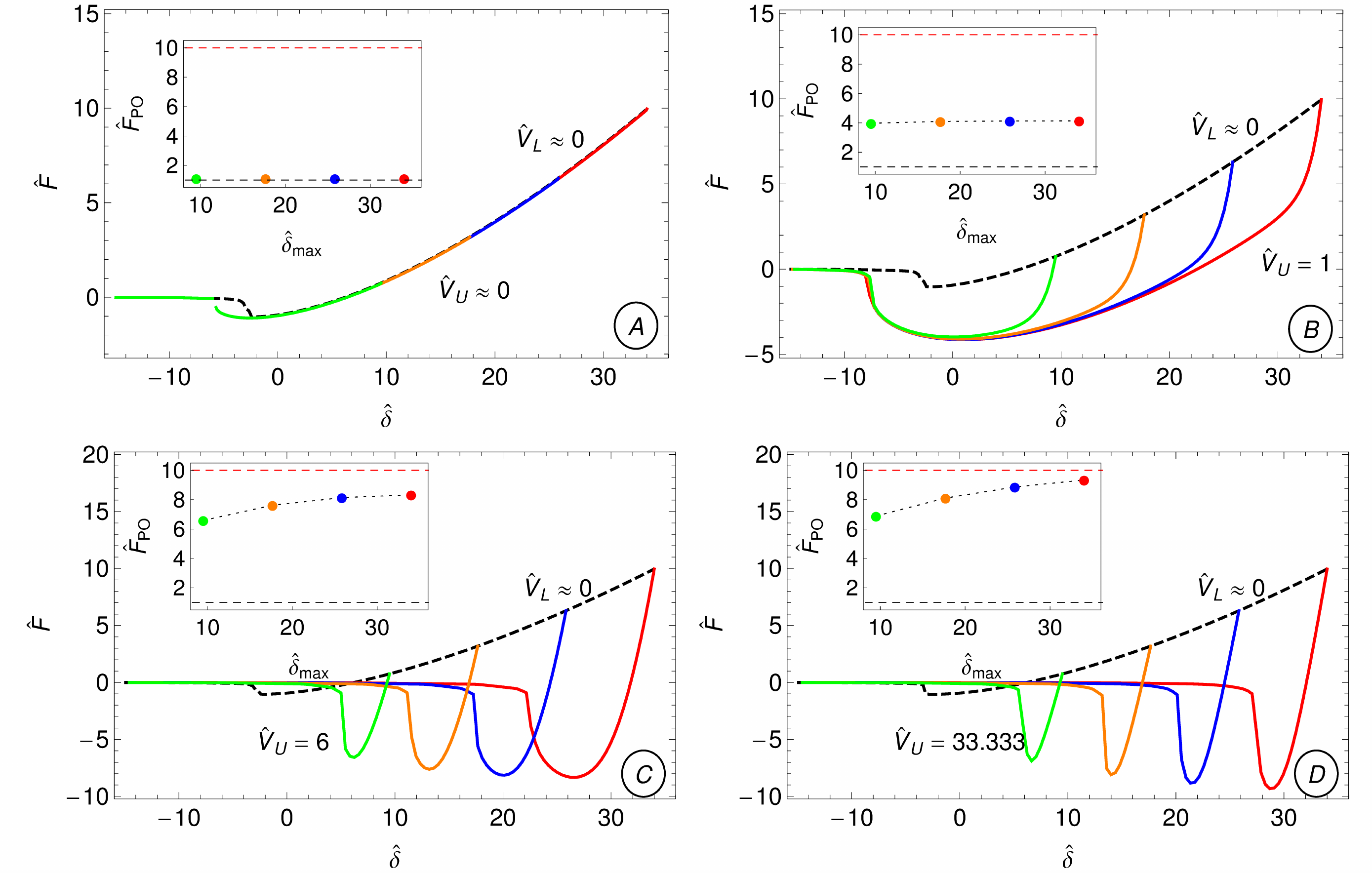}
\end{center}
\caption{ The force $\hat{F}$ as a function of the approach $\hat{\protect%
\delta}$. Loading (black dashed line) is performed at vanishing speed ($\hat{%
V}_{L}\approx 0$). Unloading (coloured solid lines) is instead performed at
different pulling speeds $\hat{V}_{U}=0,$ $1,$ $6,$ $33.333$ and from
different maximum penetrations $\hat{\protect\delta}_{\max }=9.5,$ $17.7,$ $%
25.8,$ $34$ (corresponding to $\hat{a}_{\max }=100,$ $125,$ $140,$ $160$).
In the inset of each subfigure, the normalized pull-off force $\hat{F}_{%
\mathrm{PO}}=F_{\mathrm{PO}}/(1.5\protect\pi \Delta \protect\gamma R)$ is
plotted in terms of $\hat{\protect\delta}_{\max }$; black and red dashed
lines represent JKR "elastic" limit and the upper-bound limit of eq. (%
\protect\ref{maxFPO}), respectively. Results are given for $\hat{R}=500$.}
\label{FIGURE3}
\end{figure}

Figure \ref{FIGURE3} shows the applied force $\hat{F}$ in terms of the
penetration $\hat{\delta}$, for different unloading speeds $\hat{V}$ and
maximum penetrations $\hat{\delta}_{\max }$. For vanishing unloading
velocity (Fig.\textbf{\ }\ref{FIGURE3}A), the pull-off force is independent
of $\hat{\delta}_{\max }$, as shown in the inset of the figure. Similarly,
the hysteresis loss, which is given by the area between loading-unloading
paths, is not affected by the point from which unloading starts. These
results are expected as for small $V$ the material response is elastic and
falls in the rubbery region. In such case, "elastic" adhesion hysteresis is
due to the different values of penetration at which jump-in and jump-off
instabilities occur \cite{Greenwood1997}. For unloading velocity $\hat{V}=1$ (Fig.\textbf{\ }\ref{FIGURE3}B), viscoelastic effects occur and, although $F_{\mathrm{PO}}$ is almost independent of the maximum penetration, hysteresis loss clearly increases with $\hat{\delta}_{\max }$. On the other hand, for higher unloading velocities (Figs.\textbf{\ }\ref{FIGURE3}C-D),
both the pull-off force and hysteresis loss grow with $\hat{\delta}_{\max }$. Interestingly, when $\hat{\delta}_{\max }$ is increased, pull-off occurs
at higher penetrations, which are positive (compressive) at high speeds. In
fact, at high $\hat{V}$, the sphere imprint on the substrate is still
observed even when the sphere is completely detached \cite{Afferrante2021}.

\begin{figure}[tbp]
\begin{center}
\includegraphics[width=16.0cm]{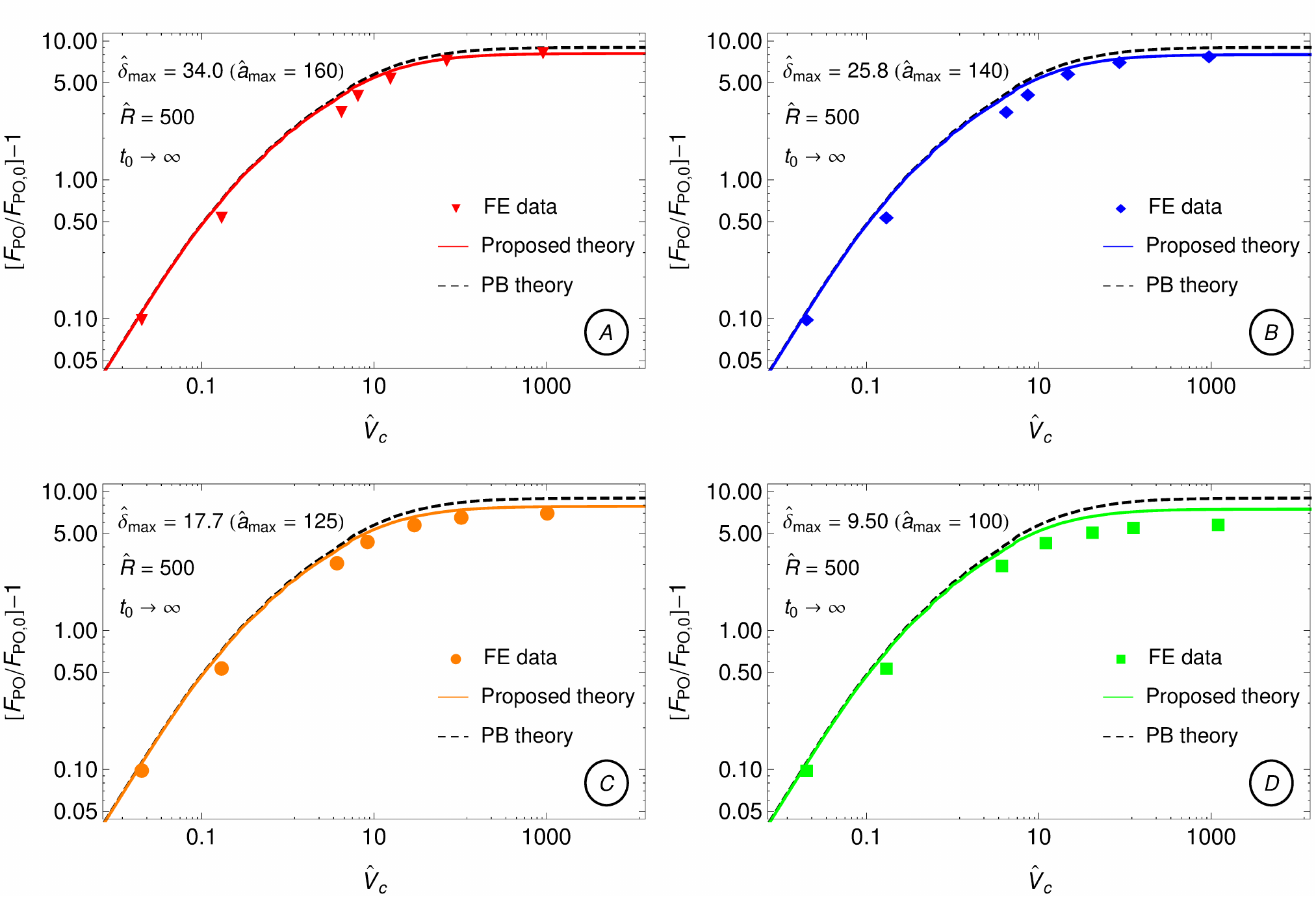}
\end{center}
\caption{ The relative increase of the pull-off force $F_{\mathrm{PO}}/F_{%
\mathrm{PO(}\hat{V}\mathrm{_{\mathrm{c}}\sim 0)}}-1$ in terms of the contact
line speed $\hat{V}_{\mathrm{c}}$, being $F_{\mathrm{PO(}\hat{V}\mathrm{_{%
\mathrm{c}}\sim 0)}}=1.5\protect\pi \Delta \protect\gamma R$. Markers refer
to FE data; black dashed line refers to PB theory \protect\cite{PB2005} for
systems of infinite size; coloured solid lines refer to the proposed eq. (%
\protect\ref{PBnew}). Results are given for $\hat{R}=500$ and different $%
\hat{a}_{\max }$. We assumed $s_{0}=1.5$ \textrm{nm} in the theory.}
\label{FIGURE4}
\end{figure}

Figure \ref{FIGURE4} shows the relative increase in viscoelastic pull-off
force with respect to the elastic one $F_{\mathrm{PO,0}}=1.5\pi \Delta
\gamma R$. Data are given in terms of the contact line velocity $\hat{V}_{%
\mathrm{c}}$ calculated at the pull-off and are compared with the
theoretical predictions of eq. (\ref{PBnew}) (coloured
solid lines). PB theory for systems of infinite size is plotted with black
dashed line. Size-effects related to $a_{\max }$ entail the
ratio $F_{\mathrm{PO}}/F_{\mathrm{PO,0}}$ reaches values lower than $%
E_{\infty }/E_{0}=10$, which is instead approached for infinite system.

\begin{figure}[tbp]
\begin{center}
\includegraphics[width=8.0cm]{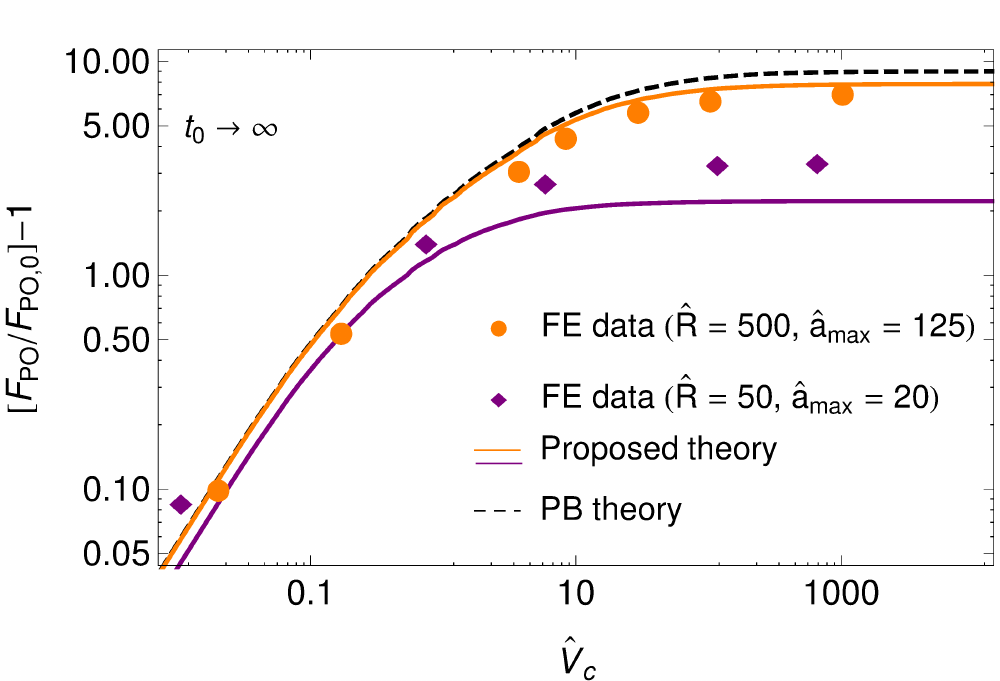}
\end{center}
\caption{ The relative increase of the pull-off force $F_{\mathrm{PO}}/F_{%
\mathrm{PO(}\hat{V}\mathrm{_{\mathrm{c}}\sim 0)}}-1$ in terms of the contact
line speed $\hat{V}_{\mathrm{c}}$, being $F_{\mathrm{PO(}\hat{V}\mathrm{_{%
\mathrm{c}}\sim 0)}}=1.5\protect\pi \Delta \protect\gamma R$. Markers refer
to FE data; black dashed line refers to PB theory \protect\cite{PB2005} for
systems of infinite size; coloured solid lines refer to the proposed eq. (%
\protect\ref{PBnew}). Results are given for $\hat{R}=50$ and $\hat{a}_{\max
}=20$. We assumed $s_{0}=1.5$ \textrm{nm} in the theory.}
\label{FIGURE5}
\end{figure}

\begin{figure}[tbp]
\begin{center}
\includegraphics[width=16.0cm]{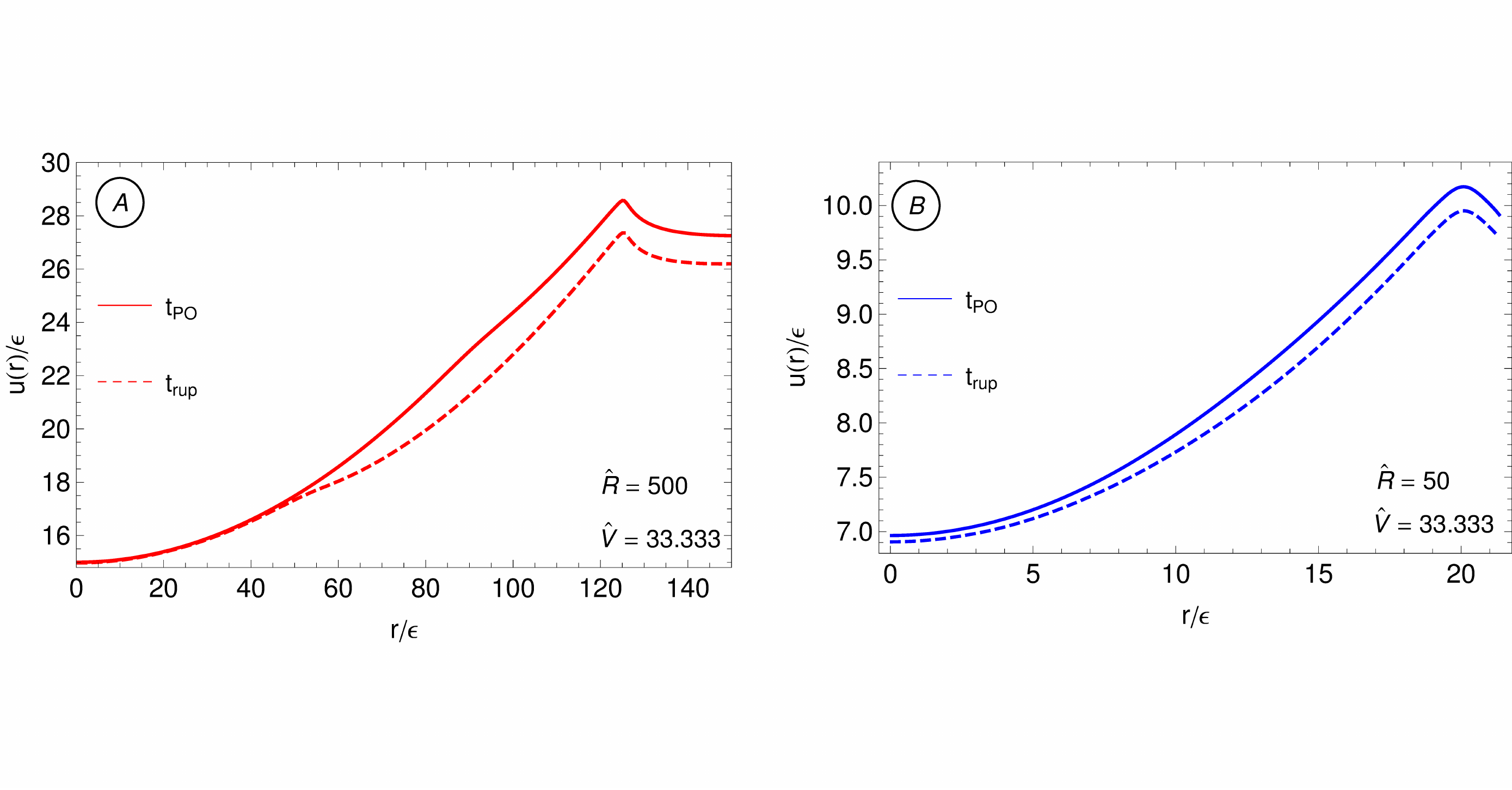}
\end{center}
\caption{Displacement fields when the pull-off force is reached ($t_{\mathrm{%
PO}}$), and at the moment of final rupture ($t_{\mathrm{rup}}$) for a
pulling velocity $\hat{V}_{U}=33.333$ and radii of curvature $\hat{R}=500$
(A) and $\hat{R}=50$ (B).}
\label{FIGURE6}
\end{figure}

Figure \ref{FIGURE5} shows that $F_{\mathrm{PO(\max )}}/F_{\mathrm{PO,0}}$ noticeably reduces when a lower radius $\hat{R}=50$ is considered. In general, numerical data and theoretical predictions are almost in agreement for $\hat{R}=500$, while some difference is observed for $\hat{R}=50$, where the theory underestimates the pull-off force. As discussed in Ref. \cite%
{Muser2021}, such differences may be due to a transition of the detachment
mode from crack propagation to quasi-uniform bond breaking, which is
expected at small scales \cite{Gao2004,PerssonNANO}. This finding is
confirmed in Fig. \ref{FIGURE6} showing the displacement fields. For $\hat{R}%
=500$, moving from the time at which the tensile force is maximun ($t_{%
\mathrm{PO}}$) to the time of final rupture ($t_{\mathrm{rup}}$), crack
propagation is clearly the mechanism of debonding; for $\hat{R}=50$, the
displacement field instead moves homogeneously during rupture.

\begin{figure}[tbp]
\begin{center}
\includegraphics[width=8.0cm]{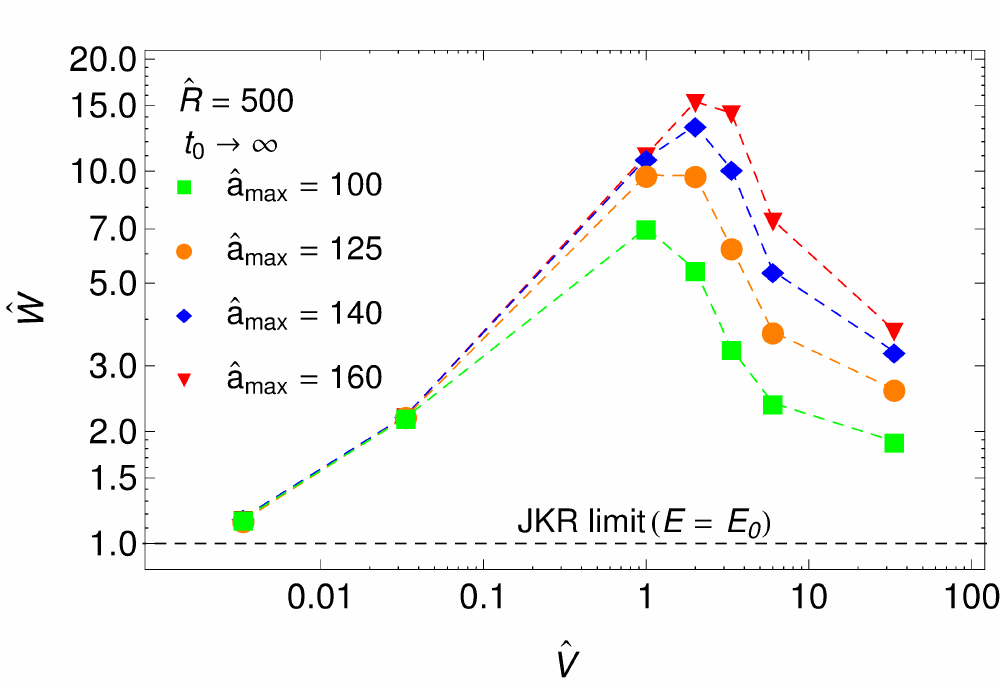}
\end{center}
\caption{ The work of separation $W$ normalized with respect to JKR value,
as a function of the pulling speed $\hat{V}$. Markers refer to FE data;
black dashed line refers to the "elastic" quasi-static JKR\ limit for $E(%
\protect\omega )=E_{0}$. Results are given for $\hat{R}=500$ and different $%
\hat{a}_{\max }$.}
\label{FIGURE7}
\end{figure}

Figure \ref{FIGURE7}\textbf{\ }shows the work of separation in terms of the
normalized pulling speed $\hat{V}$. Being simulations performed under
displacement controlled conditions, $W$ is calculated as $W=\int_{\delta _{%
\mathrm{snap-off}}}^{\delta _{\mathrm{(}F=0\mathrm{)}}}F(\delta )d\delta$,
where $\delta _{\mathrm{snap-off}}$ is the contact penetration at which
snap-off occurs and $\delta _{\mathrm{(}F=0\mathrm{)}}$ is the the
penetration corresponding to zero applied load. Results are normalized with
respect to the value calculated in the elastic limit (JKR limit). Unlike the pull-off force $F_{\mathrm{PO}}$ that is a monotonically increasing function of $V\mathrm{_{\mathrm{c}}}$, $W$ tends to the JKR limit at vanishing velocities and reaches a maximum at intermediate speeds. This may appear counter-intuitive, but the reduction in penetration is small at high pulling velocities $V$ (i.e., when the tensile force takes the highest values) due to a "stick" effect in the initial phase of debonding \cite{ViolanoFRONT, Haiat2003}; moreover, such stick effect increases with $\delta _{\max }$. In addition, once the pull-off point is passed, $V\mathrm{_{\mathrm{c}}}$ tends to quickly increase especially before snap-off and the assumption of crack-tip velocity slowly changing is no longer satisfied. This is the reason for which the work of separation remains higher than the JKR limit at high $V$ \cite{Muser2021}.

\subsection{Unloading from unrelaxed state}

A second set of simulations has been run with approach and retraction
performed at the same driving speed $V$ and with zero dwell time ($t_{0}=0$)
between the loading and unloading phases. As a result, the approximation of
quasi-static loading is no longer valid, and time dependent effects cannot
be neglected. Furthermore, being $t_{0}=0$, the stresses in the substrate
have no time to relax.

\begin{figure}[tbp]
\begin{center}
\includegraphics[width=16.0cm]{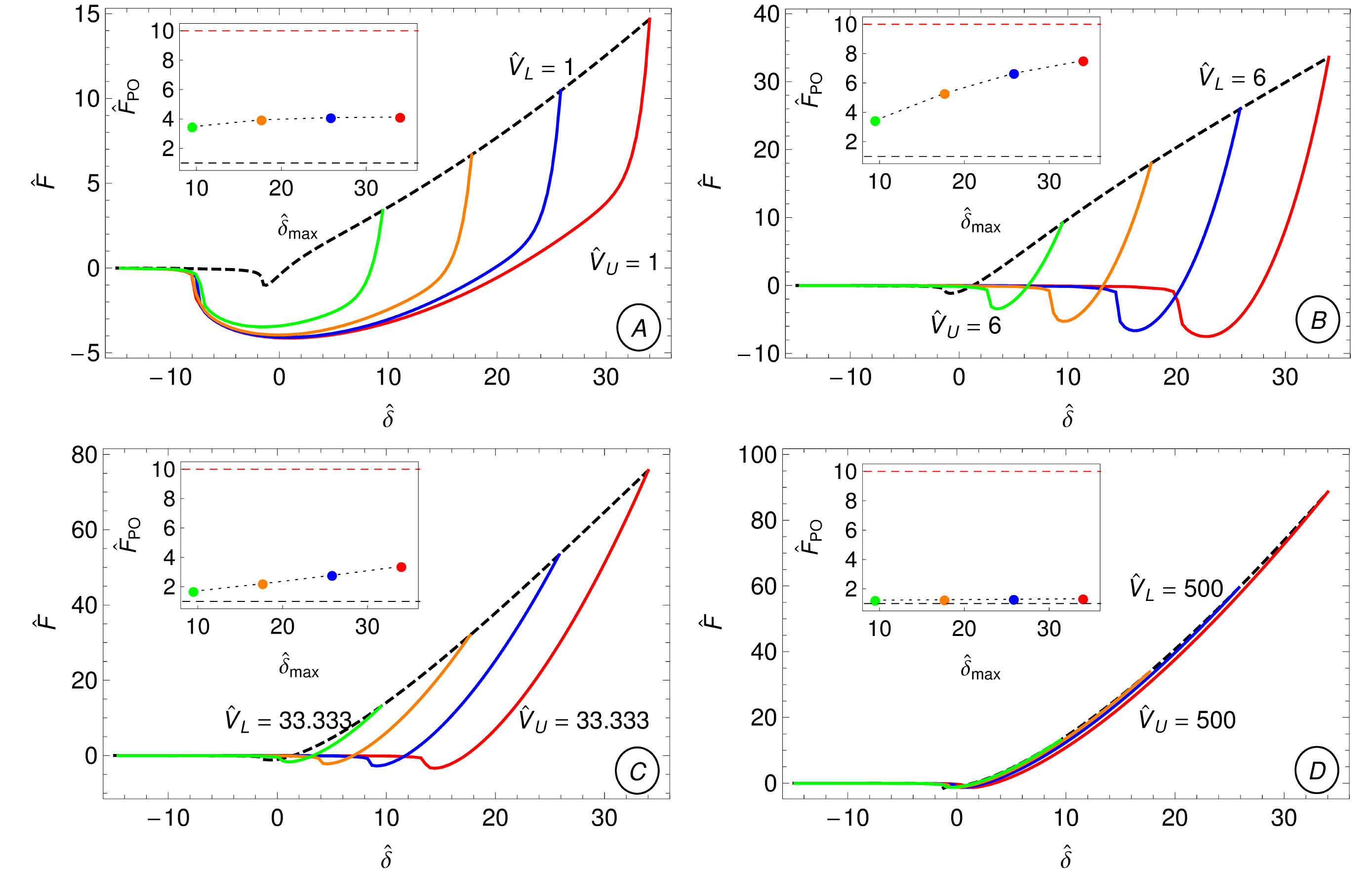}
\end{center}
\caption{ The force $\hat{F}$ as a function of the imposed approach $\hat{%
\protect\delta}$. Loading (black dashed line) and unloading (coloured solid
lines) are performed at the same nonvanishing velocity ($\hat{V}_{L}=\hat{V}%
_{U}=$ $1,$ $6,$ $33.333,$ $500$). Unloading starts from different maximum
penetrations $\hat{\protect\delta}_{\max }=9.5,$ $17.7,$ $25.8,$ $34$
(corresponding to $\hat{a}_{\max }\sim 100,$ $125,$ $140,$ $160$). In the
inset of each subfigure, the normalized pull-off force $\hat{F}_{\mathrm{PO}%
}=F_{\mathrm{PO}}/(1.5\protect\pi \Delta \protect\gamma R)$ is plotted in
terms of $\hat{\protect\delta}_{\max }$; black and red dashed lines
represent the JKR "elastic" limit and the upper-bound limit of eq. (\protect
\ref{maxFPO}), respectively. Results are given for $\hat{R}=500$.}
\label{FIGURE8}
\end{figure}

Figures \ref{FIGURE8}A-D show the force $\hat{F}$ in terms of the
penetration $\hat{\delta}$ at various loading-unloading speeds $\hat{V}=1,$ $%
6,$ $33.333,$ $500$ and different maximum penetrations $\hat{\delta}_{\max
}=9.5,$ $17.7,$ $25.8,$ $34$. During approach, viscous effects lead to a
reduction in the effective surface energy, as the system behaves similarly to a
closing circular crack \cite{Greenwood2004}. As a result, the loading path
is completely different from that observed when $V\approx 0$ (Figs. \ref%
{FIGURE3}A-D). Moreover, the increase in the pull-off force with $\delta
_{\max }$ is more marked already at relatively small velocities. It is
interesting to observe what happens at very high loading-unloading rate (Fig.%
\textbf{\ }\ref{FIGURE8}D); in this case, the viscoelastic material is
excited in its glassy region and behaves as a stiff elastic medium with
modulus $E(\omega )\approx E_{\infty }$. Consequently, the pull-off force is
found to be less dependent on $\delta_{\max }$ and closer to the JKR\
limit $F_{\mathrm{PO,0}}=1.5\pi \Delta \gamma R$ (which is independent of
the value of $E$).

\begin{figure}[tbp]
\begin{center}
\includegraphics[width=16.0cm]{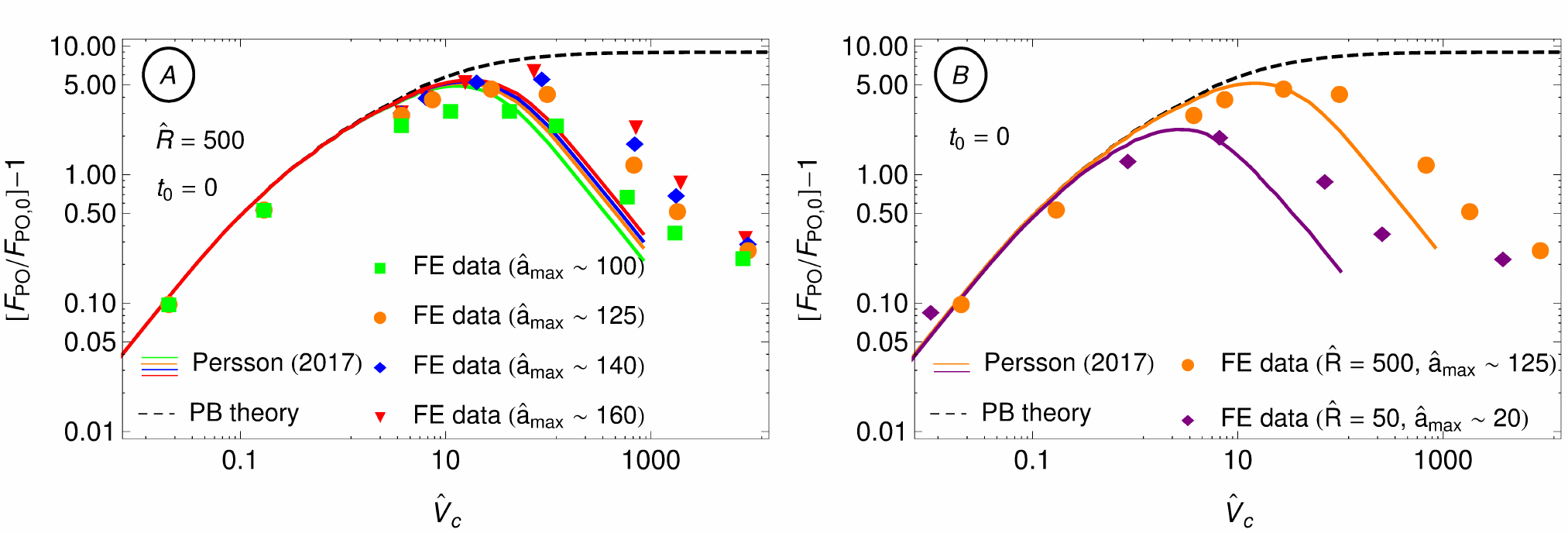}
\end{center}
\caption{The relative increase of the pull-off force $F_{\mathrm{PO}}/F_{%
\mathrm{PO(}\hat{V}\mathrm{_{\mathrm{c}}\sim 0)}}-1$ in terms of the contact
line velocity $\hat{V}_{\mathrm{c}}$, being $F_{\mathrm{PO(}\hat{V}\mathrm{_{%
\mathrm{c}}\sim 0)}}=1.5\protect\pi \Delta \protect\gamma R$. Markers refer
to FE data; black dashed line refers to PB theory \protect\cite{PB2005} for
systems of infinite size; coloured solid lines refer to Persson's theory 
\protect\cite{Persson2017}, where in eq. (\protect\ref{PBnew2}) the
integrals are calculated between $\protect\omega _{\mathrm{L}}=\protect%
\kappa 2\protect\pi V_{\mathrm{c}}/a_{\max }$ and $\protect\omega (s)=2%
\protect\pi V_{\mathrm{c}}/s$. We assumed $s_{0}=1.5$ \textrm{nm} in the
theory. A) Results are given for $\hat{R}=500$ ($\protect\kappa =1/20$) and
different $\hat{a}_{\max }$. B) Results are given for $\hat{R}=50$ ($\protect%
\kappa =1/10$), and $\hat{R}=500$ ($\protect\kappa =1/20$).}
\label{FIGURE9}
\end{figure}

Figures \ref{FIGURE9}A-B shows the relative increase in the viscoelastic
pull-off force as a function of the contact line velocity $V_{\mathrm{c}}$
(calculated at the pull-off). In this case, the maximum value of $F_{\mathrm{%
PO}}$ does not occur for high $V_{\mathrm{c}}$, but a bell-shape curve is
obtained in a double logarithmic representation. Moreover, the peak
increases with $a_{\max }$ and occurs at higher $V_{\mathrm{c}}$ when
unloading starts from higher contact radius.

When unloading starts from an unrelaxed state of the substrate, we have shown in Ref. \cite{Afferrante2021} that viscous dissipation may involve the bulk material. Therefore, the frequency of excitation can be estimated as $\omega \approx V/a$, being $V$ the normal pulling velocity \cite{Lorenz2013}. At pull-off, we have $V_{\mathrm{c}}/V\approx 20$ when $\hat{R}=500$ and speeds sufficiently high ($V\geq 6$). On the contrary, for $\hat{R}=50$ we find $V_{\mathrm{c}}/V\approx 10$ in the limit of high speeds. As in eq. (\ref{PBnew2}), the cut-off frequency is estimated in terms of $V_{\mathrm{c}}$, we can reasonably assume $\omega _{\mathrm{L}}=\kappa 2\pi V_{\mathrm{c}}/a_{\max }$, with $\kappa \sim 1/10$ for $\hat{R}=50$ and $\kappa \sim 1/20$ for $\hat{R}=500$. The resulting
theoretical predictions are compared with numerical data in Figs. \ref{FIGURE9}A-B, where the solution for systems of infinite size is plotted
with black dashed line. We observe that the maximum contact radius $a_{\max }$ increases with the loading speed $V$ in numerical simulations but this effect cannot be considered in theoretical calculations. For this reason, in evaluating theoretical curves always the same value of $a_{\max }$ calculated at $V\approx 0$ is considered after we verified it negligibly affects Persson's curves.

\begin{figure}[tbp]
\begin{center}
\includegraphics[width=8.0cm]{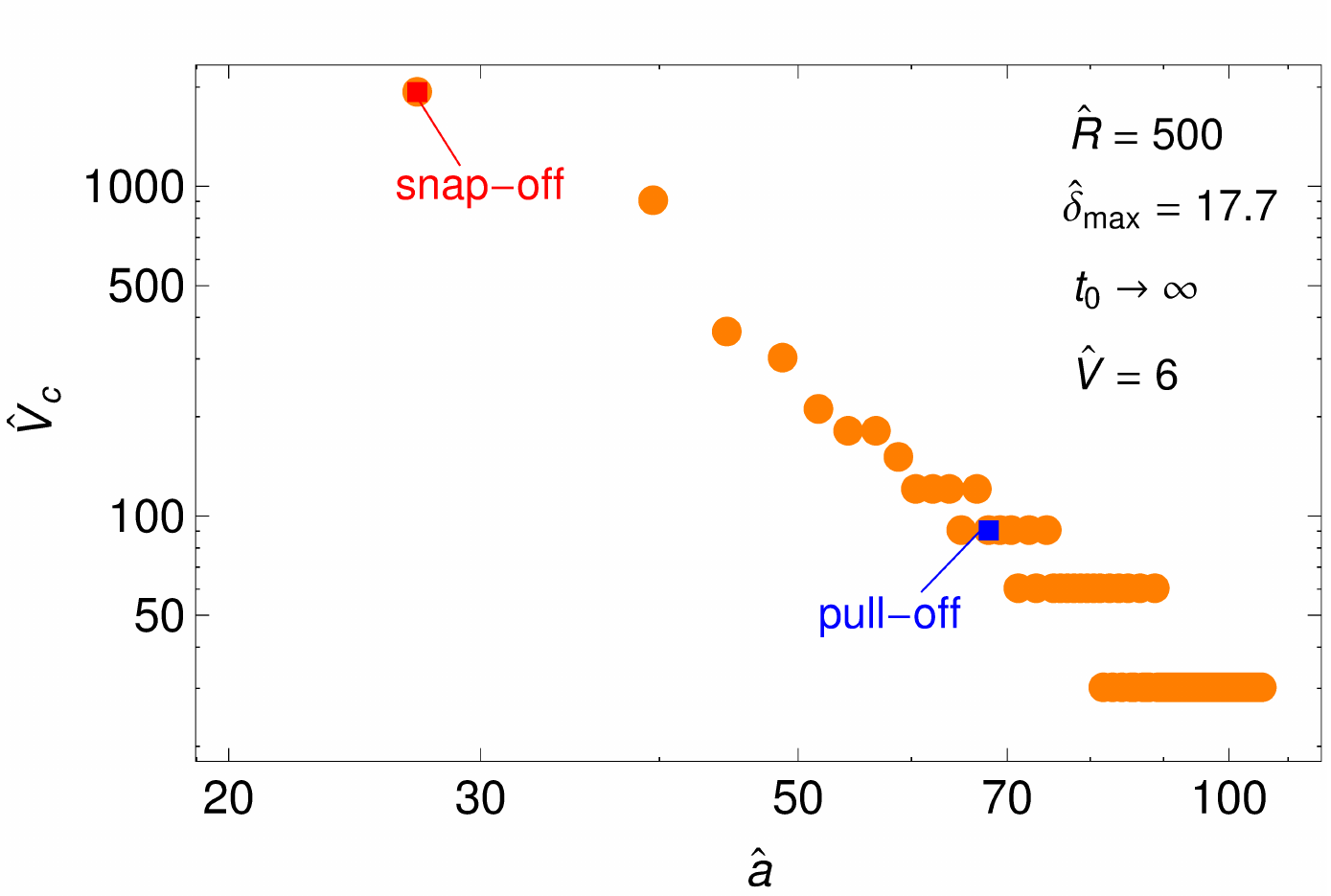}
\end{center}
\caption{The normalized contact line velocity $\hat{V}_{\mathrm{c}}$ in
terms of the contact radius $\hat{a}$, when unloading at a pulling speed $%
\hat{V}=6$.}
\label{FIGURE10}
\end{figure}

Theory and numerical data are in qualitative agreement even if differences
occur in the range of high velocities. Such differences can be explained by
observing that the theory returns the effective surface energy "\textit{assuming that in the absence of adhesion, there is no elastic energy left in
the viscoelastic solid after removing the spherical indenter}" \cite{Persson2021}, while some elastic energy will be necessarily left after
unloading. Furthermore, Persson's theory assumes that the gross of
viscous dissipation is localized at the crack tip while, in this case, dissipation also involves bulk material \cite{Afferrante2021}. In addition, we remark
that the theory assumes steady-state crack propagation, while in our
simulations $V_{\mathrm{c}}$ is not constant during detachment as
shown in Fig. \ref{FIGURE10} where the dependence of $\hat{V}_{\mathrm{c}}$ on the contact radius $\hat{a}$\ is plotted for a pulling
speed $\hat{V}=6$. The contact line velocity $\hat{V}_{\mathrm{c}}$
varies widely (about two orders of magnitude) moving from $\hat{a}_{\max}$ to
snap-off, and the pull-off point is not reached under steady-state
conditions. 

\begin{figure}[tbp]
\begin{center}
\includegraphics[width=8.0cm]{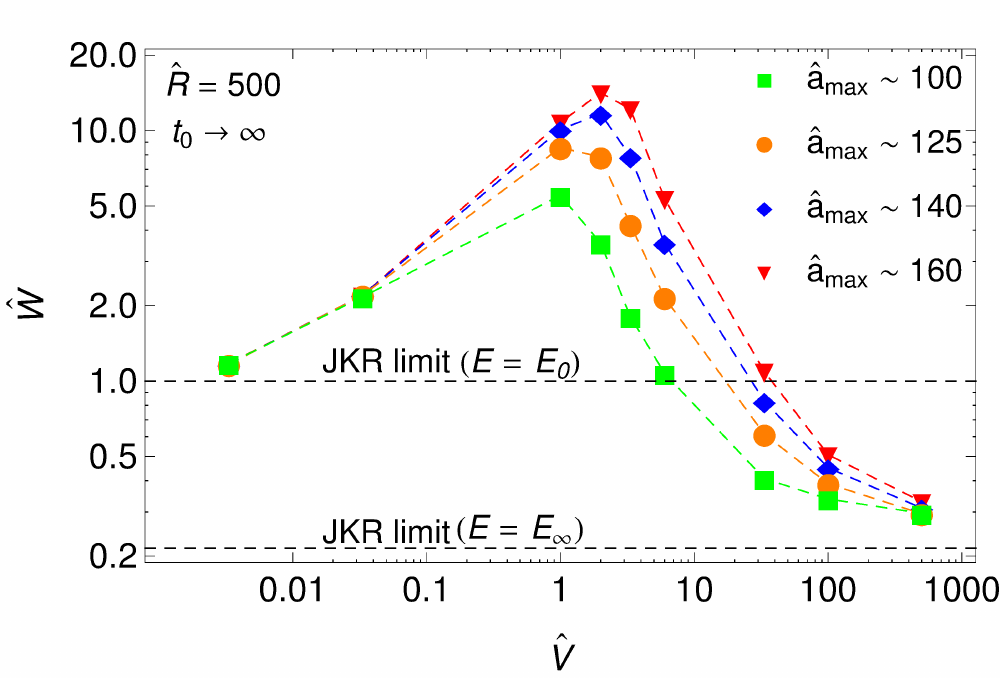}
\end{center}
\caption{The normalized work of separation $\hat{W}$ in terms of the pulling
speed $\hat{V}$. Markers refer to FE data; black dashed lines refer to the
"elastic" quasi-static JKR\ limit for $E(\protect\omega )=E_{0}$ and $E(%
\protect\omega )=E_{\infty }$. Results are given for $\hat{R}=500$ and
different $\hat{a}_{\max }$.}
\label{FIGURE11}
\end{figure}

Finally, Fig. \ref{FIGURE11}\textbf{\ }shows the work of separation in terms
of the pulling velocity for different initial values of the contact radius. $W$
increases with $a_{\max }$ except at low and high speeds where the
viscoelastic substrate responds elastically with $E(\omega )=E_{0}$ at
vanishing speeds, and $E(\omega )=E_{\infty }$ at very high $V$. In such
limits, the JKR values of $W$ are recovered.

\section{Conclusions}

Size-dependent effects in the adhesion of soft materials find partial
explanation in the scientific literature, where the correlation between size
and rate effects is often neglected. For this reason, in this work, we try
to shed some light on this problem through the investigation of the adhesive
contact between a rigid sphere and a viscoelastic substrate, by exploiting a
recent FE model developed in Ref. \cite{Afferrante2021}.

We have performed various simulations under different loading and unloading
conditions. When unloading starts from a fully relaxed state of the
viscoelastic material, the pull-off force (and hence the effective surface
energy) is a monotonic increasing function of the contact line velocity $V_{%
\mathrm{c}}$. For systems of infinite size, Persson \& Brener (PB) \cite%
{PB2005} theory predicts $F_{\mathrm{PO(\max )}}/F_{\mathrm{PO,0}}=E_{\infty
}/E_{0}$ at high $V_{\mathrm{c}}$, while numerical calculations show that
the asymptotic maximum value $F_{\mathrm{PO(\max )}}$\ of the pull-off force
is affected by the finite dimension of the contact radius. For this reason,
we suggest introducing a corrective factor $\alpha (s_{0},a_{\max })$ in PB
theory which takes into account geometric effects related to the system
under investigation (and, in particular, the finite value of the contact
radius $a_{\max }$ reached at the end of loading). A good agreement is found
between theory and numerical data, although some differences are observed
for the smaller curvature radii of the sphere, as a transition of the
detachment mode from crack propagation to quasi-uniform bond-breaking may
occur at small scales.

When loading-unloading are performed at the same nonvanishing driving speed $%
V$ (and without waiting time between approach and retraction), the pull-off
force $F_{\mathrm{PO}}$ shows a peak at intermediate $V_{\mathrm{c}}$. Such
peak increases with $a_{\max }$ and moves towards higher $V_{\mathrm{c}}$
when unloading starts from a higher initial contact radius. At very high $V_{%
\mathrm{c}}$, the substrate behaves again elastically with an elastic
modulus $E_{\infty }$. Also, Persson's theory \cite{Persson2017,Persson2021} agrees with FE calculations but some quantitative differences can be observed at the highest speeds as a result of the simplifying assumptions included in the theoretical model.

\section*{ACKNOWLEDGEMENTS}
G.V. and L.A. acknowledge support from the Italian Ministry of Education, University and Research (MIUR) under the programme
"Departments of Excellence" (L.232/2016).

\end{document}